\documentclass[12pt]{article}
\usepackage{graphicx}

\def\valpo{$^{1}$Departamento de Fisica, Universidad Tecnica Federico Santa Maria; \\
Centro Cientifico-Tecnologico de Valparaiso, Casilla 110-V, Valparaiso, Chile }
\def\prague{$^{2}$Czech Technical University in Prague, FNSPE, Brehova 7, 11519 Prague, Czech Republic  }
\def\Title#1{\begin{center} {\Large #1 } \end{center}}
\def\Author#1{\begin{center}{ \sc #1} \end{center}}
\def\Address#1{\begin{center}{ \it #1} \end{center}}

\newenvironment{Abstract}{\begin{quotation}  }{\end{quotation}}
\newenvironment{Presented}{\begin{quotation} \begin{center}
             PRESENTED AT\end{center}\bigskip
      \begin{center}\begin{large}}{\end{large}\end{center} \end{quotation}}
\def\Acknowledgements{\bigskip  \bigskip \begin{center} \begin{large}
             \bf ACKNOWLEDGEMENTS \end{large}\end{center}}
\begin{document}
\begin{titlepage}
%\pubblock

\vfill
\Title{Spin Dependence of Small-Angle Proton-Nucleus Scattering }
\vfill
\Author{Michal Krelina$^{1,2}$ and Boris Kopeliovich$^{1}$}
\Address{\valpo\\
\prague}
\vfill
\begin{Abstract}
We study the single-spin asymmetry, $A_N(t)$, arising from Coulomb-nuclear interference (CNI) at small 4-momentum transfer squared, $-t=q^2$, aiming at explanation of the recent data from the PHENIX experiment at RHIC on polarized proton-nucleus scattering, exposing a nontrivial $t$-dependence of $A_N$.
We found that the failure of previous theoretical attempts to explain these data, was due to
lack of absorptive corrections in the Coulomb amplitude of $pA$ elastic scattering.
Our prominent observation is that the main contribution to $A_N(t)$ comes from interference of the amplitudes of ultra-peripheral and central collisions.
\end{Abstract}
\vfill
\begin{Presented}
EDS Blois 2017 \\
Prague, Czech Republic, June 26-30, 2017
\end{Presented}
\vfill
\end{titlepage}
\def\thefootnote{\fnsymbol{footnote}}
\setcounter{footnote}{0}
%

%\section{Introduction}

Coulomb-nuclear interference (CNI) results in a sizeable single-spin asymmetry $A_N$
\cite{kl74,Kopeliovich-a},
which is nearly energy independent and offers an uniques opportunity of measuring the spin-flip component of the Pomeron \cite{kz}.
However,  data for CNI generated $A_N$ are available so far only from fixed target experiments \cite{Poblaguev}, in the energy range, not high enough to completely neglect the contribution of Reggeons with large spin-flip part. The effective way to suppress iso-vector Reggeons, which have maximal spin-flip components, is to use nuclear targets, which either completely eliminate
such Reggeons, or suppress by factor $(1-Z/A-1)$. Although the first calculations of nuclear effects \cite{Kopeliovich-b,Kopeliovich-a} well reproduced data for $A_N$ in proton-carbon elastic scattering, measurements on heavier targets, especially gold, revealed an unexpected
$t$-dependence \cite{Poblaguev}.

Therefore, it is usual to introduce a ratio of spin-flip to the imaginary part of the non-flip hadronic elastic amplitudes, $r_5$, which is difficult to extract from experimental data due to
vanishingly small phase shift between two amplitudes \cite{Buttimore}.
It was proved in \cite{Kopeliovich-b} that this ratio $r_5$ remains unchanged in proton-nucleon compared with $pp$ scattering at  high energies.

%\begin{equation}\label{eq:r5}
%  r_5^{pA}\equiv \frac{1}{2} r_5(\rho_{pA}+i),
%\end{equation}
%where $\rho$ is the real-to-imaginary ratio of the non-flip hadronic amplitude.

%First, following the theory in \cite{Buttimore} the fits of $r_5$ for proton-proton are presented in Fig.~\ref{fig:pp} with significantly small errors for the imaginary part of $r_5$. Now, the similar values of $r_5$ we could expect also for the proton-nucleus case.

\begin{figure}[ht]
\begin{minipage}[b]{.49\textwidth}
    \centering
     \includegraphics[width=\linewidth,clip]{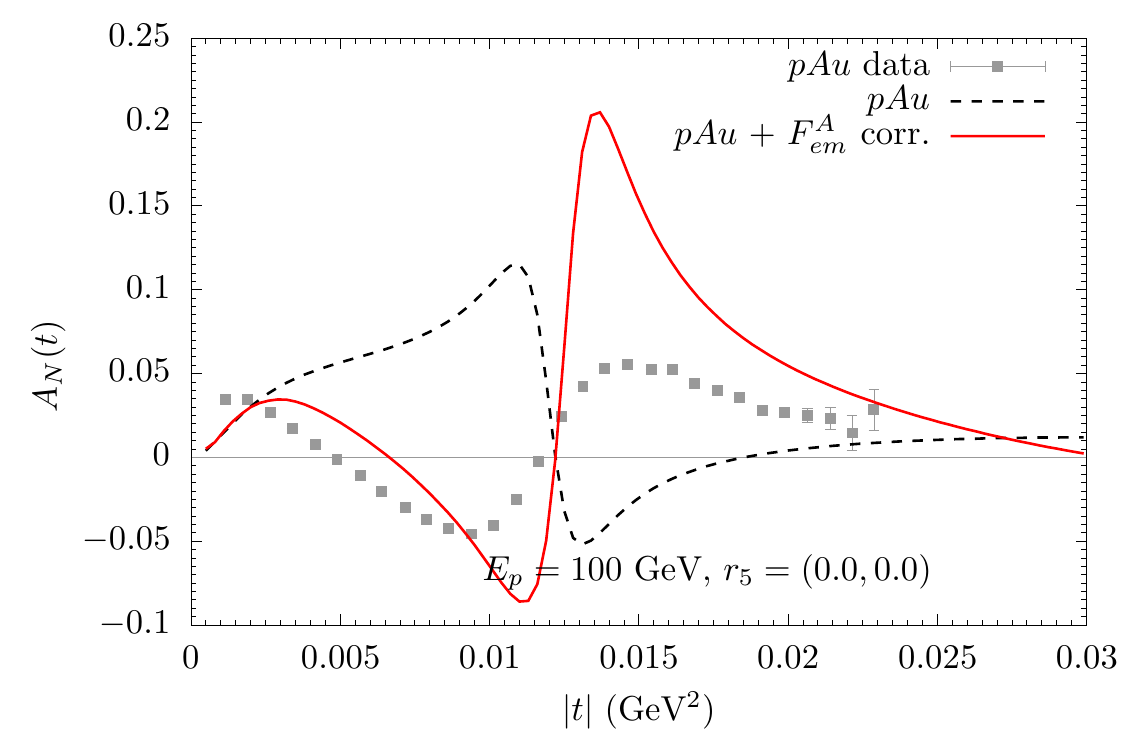}
    \caption{Power analyzing experimental data vs theory with zero $r_5$.}
    \label{fig:pAu}       % Give a unique label
\end{minipage}
\hfill
\begin{minipage}[b]{.49\textwidth}
    \centering
     \includegraphics[width=\linewidth,clip]{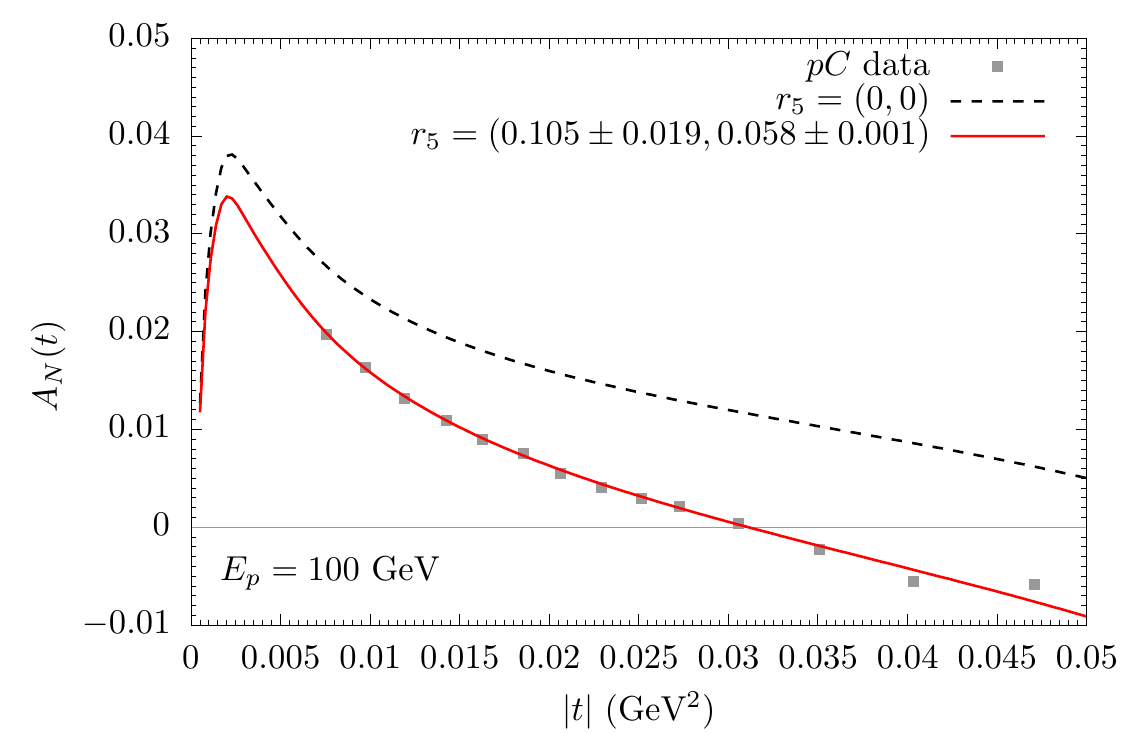}
    \caption{Power analyzing for proton-Carbon elastic scattering.}
    \label{fig:pC}       % Give a unique label
\end{minipage}
\end{figure}

However, the $t$-dependence of $A_N(t)$ in the proton-gold elastic interaction measured recently in the experiment PHENIX \cite{Poblaguev} revealed a dramatic disagreement with theoretically predicted in \cite{Kopeliovich-a}.
This is demonstrated in Fig.~\ref{fig:pAu}, where the experimental data has the nearly inverse trend in comparison with the theoretical calculations \cite{Kopeliovich-a} (dashed line, zero $r_5$). Notice, that introduction of a nonzero $r_5$ does not help to reduce the disagreement.

Here we found that the source of the trouble is the incorrect electromagnetic form factor, used in \cite{Kopeliovich-b,Kopeliovich-a}, which corresponded to the Fourier transform of the nuclear thickness function $T_A(b)=\int_{-\infty}^{\infty}dz\rho_A(b,z)$, the integral of the nuclear density along the trajectory of the incoming proton.
\begin{equation}\label{eq:Fem-old}
 F_A^{em}(t) = \frac{1}{A} \int d^2b  \,   T_A(b) e^{i \vec q \cdot \vec b},
\end{equation}
normalized as $F_A^{em}(0)=1$. Calculation with this form factor is presented in Fig.~\ref{fig:pAu} by black dashed curve ($r_5=0$).
Now, we introduce the new form of EM form factor including the absorptive corrections, which are rather strong on heavy nuclei,
\begin{equation}\label{eq:Fem-new}
  F_A^{em}(t) = \frac{1}{N^{em}} \int d^2b  \,  T_A(b) e^{i \vec q \cdot \vec b} e^{-\frac{1}{2}\sigma_{tot}^{pp}T_A(b)},
\end{equation}
where $N^{em}$ is a normalization keeping $F_A^{em}(0)=1$, and $\sigma_{tot}^{pp}$ is the total nucleon-nucleon cross section.
Therefore, the electromagnetic amplitude gets the main contribution from the ultra-peripheral collisions, $b>R_A$, while the hadronic amplitude
\begin{equation}\label{eq:Fh}
%  F_A^{h}(t) =i \int d^2b  \, e^{i \vec q \cdot \vec b}\left( 1-  e^{-\frac{1}{2}\sigma_{tot}^{pp}(1-i\rho_{pp})T_A(b)} \right)
F_A^{h}(t) =\frac{2i}{\sigma^{pA}_{tot}} \int d^2b  \, e^{i \vec q \cdot \vec b}\left( 1-  e^{-\frac{1}{2}\sigma_{tot}^{pp}T_A(b)} \right)
\end{equation}
is non-zero only at small impact parameters, $b<R_A$. In (\ref{eq:Fh}) $\sigma^{pA}_{tot}$ is the total $p+A$ cross section.
Interference of  amplitudes with different impact parameters is due to coherence of elastic scattering.
%In (\ref{eq:Fh}), $\rho_{pp}$ is the real-to-imaginary ratio of the non-flip hadronic amplitude.

Comparison of the theory with the corrected EM form factor (\ref{eq:Fem-new}) (red solid line) with experimental data presented in Fig.~\ref{fig:pAu} (zero $r_5$), demonstrates similar shapes of the $t$-dependences.

\begin{figure}[ht]
\begin{minipage}[b]{.49\textwidth}
    \centering
     \includegraphics[width=\linewidth,clip]{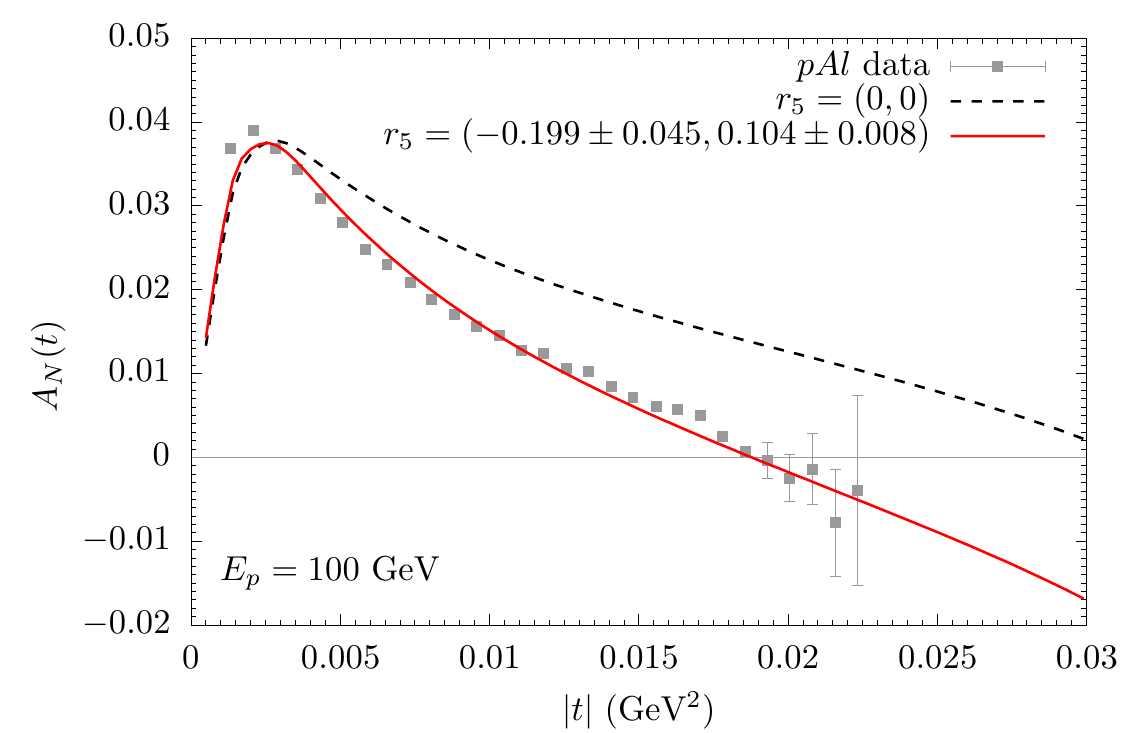}
    \caption{Power analyzing for proton-Aluminium elastic scattering.}
    \label{fig:pAl}       % Give a unique label
\end{minipage}
\hfill
\begin{minipage}[b]{.49\textwidth}
    \centering
     \includegraphics[width=\linewidth,clip]{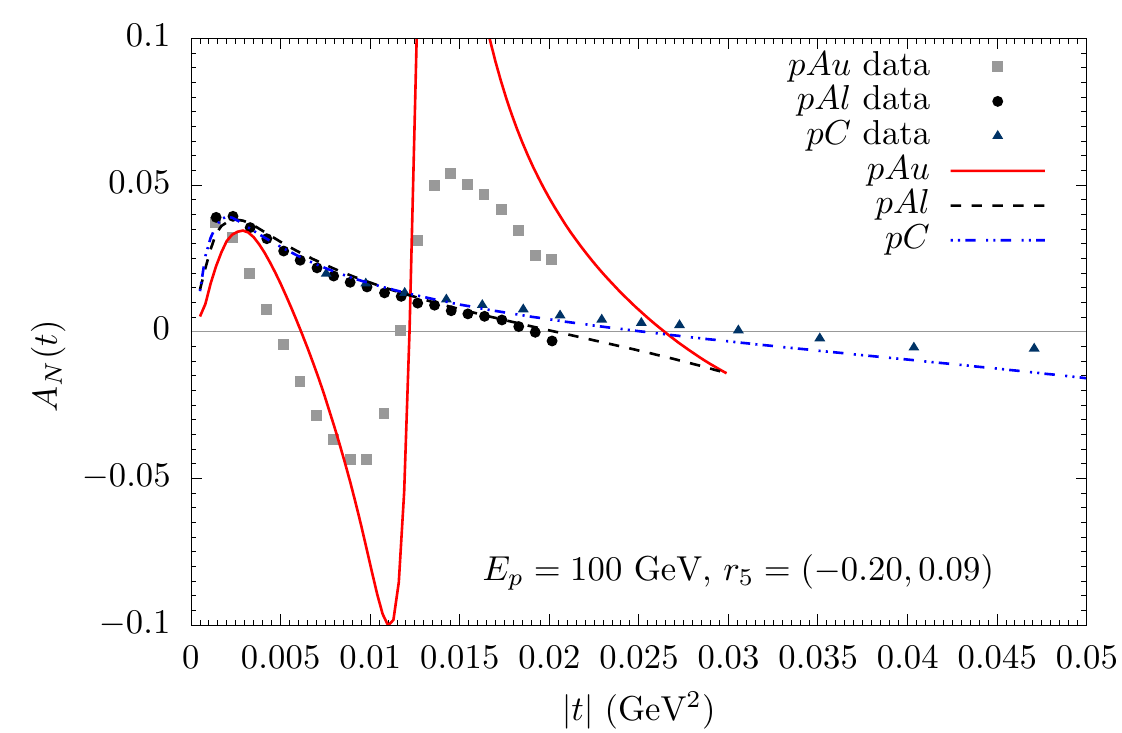}
    \caption{Experimental data vs predictions with the same $r_5$.}
    \label{fig:all}       % Give a unique label
\end{minipage}
\end{figure}

Next, using of the new EM form factor we did calculations also for other nuclei,  $p+C$ and $p+Al$ in Figs.~\ref{fig:pC} and \ref{fig:pAl}, respectively. Here, the black dashed lines correspond to $r_5=0$, and  red solid curves to the best fit of $r_5$.

Nevertheless, the values of best fits of $r_5$ for all three nuclear targets differ significantly, which is in the contradiction with our assumption about universal $r_5$.

Finally, in Fig.~\ref{fig:all} we compare all experimental data with predictions, where we used the same $r_5$ ratio for all nuclei.  The global fit gives a reasonable values for real and imaginary part of $r_5$, but not reliable so far.

In conclusion, proton-nucleus elastic scatterings in the CNI region provide an opportunity to study spin-flip hadronic interaction, where the isovector Reggeons are excluded or suppressed, and a nonzero spin-flip amplitude is required for describing experimental data.

Although we succeeded reproduce the the general features of $t$-dependence of $A_N(t)$ observed in proton-gold elastic scattering, the calculations apparently need further improvements. So far, for the sake of simplicity, we relied on the Glauber eikonal model, which it is well known to be subject to Gribov corrections \cite{gribov,kopeliovich-gribov}.
This corrections make nuclei more transparent for hadrons and may considerably affect our results. We keep working on this issue and the upcoming results will be published elsewhere.

\Acknowledgements
This work was supported in part by the Conicyt grant  PIA ACT1406 (Chile), by Fondecyt (Chile) grant 1170319, and by Proyecto Basal FB 0821 (Chile).
The work of Michal Krelina was also partially supported by the grant LTC17038 of the INTER-EXCELLENCE program at the Ministry of Education, Youth and Sports of the Czech Republic.

\end{document}